\pdfoutput=1

\documentclass[twocolumn,preprint,3p]{elsarticle}

\usepackage{amssymb}
\usepackage{lineno, hyperref}
\usepackage{multirow}
\usepackage{color}
\usepackage{tabularx, booktabs}
\usepackage{makecell}
\usepackage{tcolorbox}

\newcolumntype{Y}{>{\centering\arraybackslash}X}

\begin{document}

\onecolumn
\begin{tcolorbox}
\textbf{NOTICE}: This is the author's version of a work that was accepted for publication by Elsevier. Changes resulting from the publishing process, such as peer review, editing, corrections, structural formatting, and other quality control mechanisms may not be reflected in this document. Changes may have been made to this work since it was submitted for publication. A definitive version was subsequently published in Future Generation Computer Systems, \url{http://dx.doi.org/10.1016/j.future.2016.11.009}.
\end{tcolorbox}
\twocolumn

\begin{frontmatter}



\title{Mobile Edge Computing, Fog et al.: A Survey and Analysis of Security Threats and Challenges}


\author[es]{Rodrigo Roman}
\ead{roman@lcc.uma.es}

\author[es]{Javier Lopez}
\ead{jlm@lcc.uma.es}

\author[jp]{Masahiro Mambo}
\ead{mambo@ec.t.kanazawa-u.ac.jp}

\address[es]{Computer Science Department, University of Malaga, Ada Byron building, 29071 Malaga, Spain.}
\address[jp]{Faculty of Electrical and Computer Engineering, Institute of Science and Engineering, Kanazawa University, Kakuma Kanazawa 920-1192, Japan.}

\begin{abstract}
For various reasons, the cloud computing paradigm is unable to meet certain requirements (e.g. low latency and jitter, context awareness, mobility support) that are crucial for several applications (e.g. vehicular networks, augmented reality). To fulfil these requirements, various paradigms, such as fog computing, mobile edge computing, and mobile cloud computing, have emerged in recent years. While these edge paradigms share several features, most of the existing research is compartmentalised; no synergies have been explored. This is especially true in the field of security, where most analyses focus only on one edge paradigm, while ignoring the others. The main goal of this study is to holistically analyse the security threats, challenges, and mechanisms inherent in all edge paradigms, while highlighting potential synergies and venues of collaboration. In our results, we will show that all edge paradigms should consider the advances in other paradigms.
\end{abstract}


\end{frontmatter}


\section{Introduction}
\label{intro}

Cloud computing has taken the world by storm. In this category of utility computing, a collection of computing resources (e.g. network, servers, storage) are pooled to serve multiple consumers, using a multi-tenant model. These resources are available over a network, and accessed through standard mechanisms~\cite{NISTCC2011}. The cloud computing paradigm provides a variety of deployment models and service models, from public clouds (organizations provide cloud computing services to any customer) to private clouds (organizations deploy their own private cloud computing platform), and from Infrastructure as a Service models (IaaS, where fundamental computing resources are offered as a capability) to Software as a Service models (SaaS, where applications are offered as a capability), among other things. The benefits of cloud computing --  minimal management effort, convenience, rapid elasticity, pay per use, ubiquity -- have given birth to a multi-billion industry that is growing worldwide~\cite{IDCCC2016}.

Despite its benefits, cloud computing is not a panacea. Generally, public cloud vendors have built a few large data centers in various parts of the world. These large-scale, commodity-computer data centers have enough computing resources to serve a very large number of users. However, this centralization of resources implies a large average separation between end user devices and their clouds, which in turn increases the average network latency and jitter~\cite{Satyanarayanan15}. Because of this physical distance, cloud services are not able to directly access local contextual information, such as precise user location, local network conditions, or even information about users' mobility behaviour. For various delay-sensitive applications, such as vehicular networks and augmented reality, these requirements (low latency and jitter, context awareness, mobility support) are needed.

\begin{figure*}
\centering
\includegraphics[width=350pt]{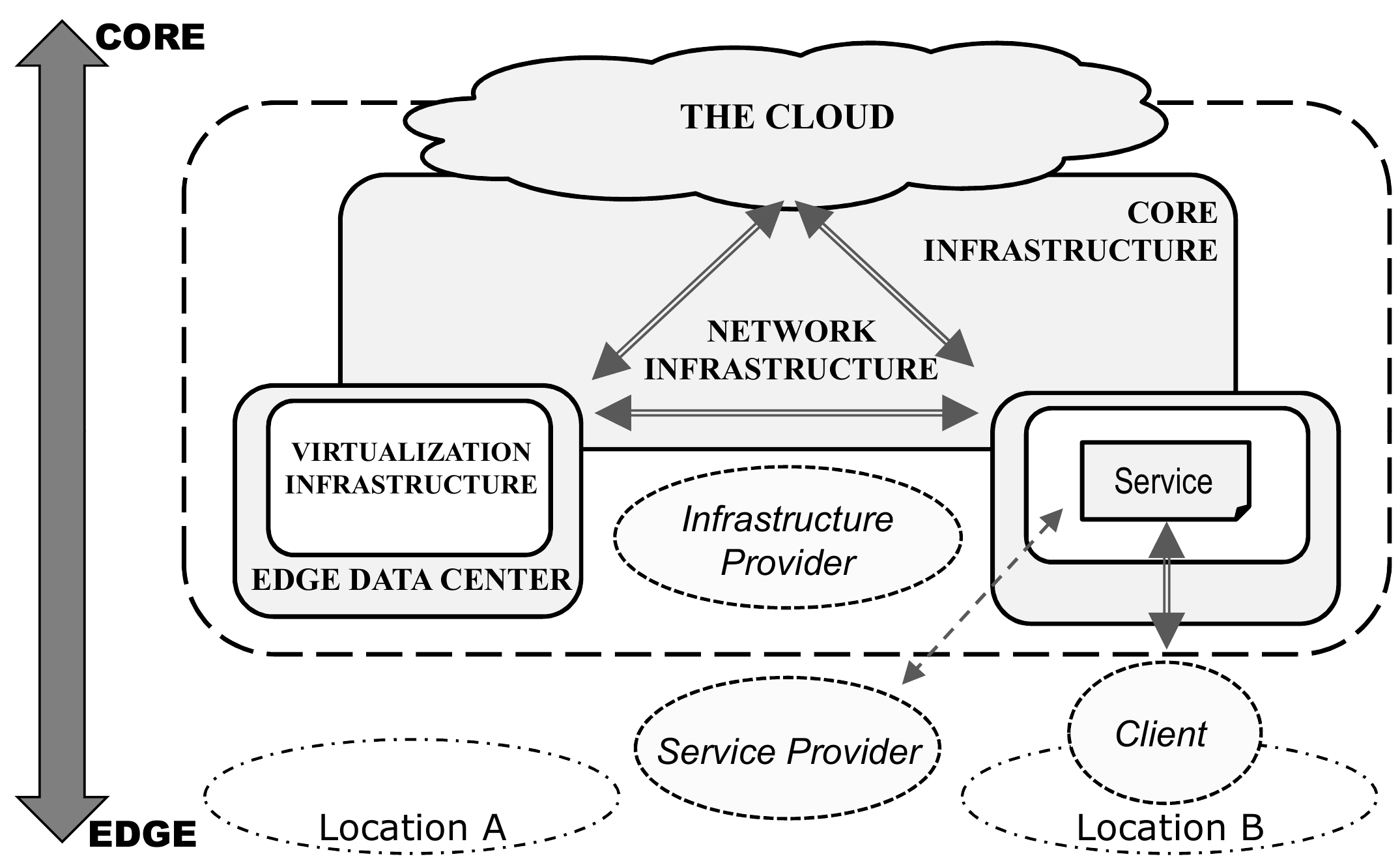}
\caption{Functional structure of edge paradigms}
\label{fig:Edge}
\end{figure*}

For these reasons, in recent years, various novel paradigms have emerged, such as fog computing~\cite{Vaquero14}, mobile edge computing~\cite{Michael14}, and mobile cloud computing ~\cite{WangMCC15}, among others (cf.~\cite{Manco2015,GarciaLopez15}). The common denominator in these edge paradigms is the deployment of cloud computing-like capabilities at the edge of the network. Most edge paradigms follow the structure shown in Figure~\ref{fig:Edge}. Edge data centers, which are owned and deployed by infrastructure providers, implement a multi-tenant virtualization infrastructure. Any customer -- from third-party service providers to end users and the infrastructure providers themselves -- can make use of these data centers' services. In addition, while edge data centers can act autonomously and cooperate with one another, they are not disconnected from the traditional cloud. It is therefore possible to create a hierarchical multi-tiered architecture, interconnected by a network infrastructure. Besides, we have to consider the potential existence of an underlying infrastructure, or core infrastructure (e.g. mobile core networks, centralized cloud services), that provide various support mechanisms, such as management platforms and user registration services. Finally, one trust domain (i.e. edge infrastructure that is owned by a infrastructure provider) can cooperate with other trust domains, creating an open ecosystem where multitude of customers can be served.

There are various differences among edge paradigms, such as the focus on mobile network operators as infrastructure providers in mobile edge computing, the existence of user-owned edge data centers (i.e. personal cloudlets) in mobile cloud computing, and the use of different underlying protocols and interfaces, among others. Nonetheless, there remain numerous similarities. Still, little of the research in these fields takes into consideration these similarities. Most architectures, protocols, services, and mechanisms are designed with only one edge paradigm in mind, and they do not consider the state of the art of other edge paradigms. At this initial stage, researchers should consider that research findings in relation to one edge paradigm might also be applied or adapted to other edge paradigms.

This silo mentality is especially conspicuous in the field of security. Although research on security issues in edge paradigms is still nascent, given the importance of this particular field, various researchers have already identified various potential threats. In the process, they have developed several security and privacy mechanisms. However, as mentioned, most research does not follow an interdisciplinary approach: studies tend to focus solely on one particular edge paradigm and its state of the art. Moreover, very few researchers have considered that it might be possible to analyse and adapt other security mechanisms that were initially designed for enabling technologies (e.g. wireless networks, distributed and peer-to-peer systems, virtualization platforms~\cite{Vaquero14}) and other related paradigms (e.g. cloud computing, grid computing).

Therefore, \textit{this study looks to provide, from a holistic perspective, a detailed analysis of the security of edge paradigms}. This analysis will be organized as follows. Section~\ref{overview} introduces the most important edge paradigms, including their history, use cases, and standardization efforts. Section~\ref{feat_syn} analyses the common features of, and differences among, all edge paradigms, and highlights both their challenges and potential synergies. Section~\ref{sec_thre} introduces the security issues that affect all edge paradigms; this section analyses the various threat models that target edge paradigms, alongside a brief overview of the requirements and challenges of the security mechanisms that should be used in this context. Section~\ref{sec_cha} presents an analysis of the current state of the art regarding security in edge paradigms. This analysis does not merely enumerate existing security mechanisms; it also points out synergies among security mechanisms originally designed for edge paradigms and other related fields. Finally, conclusions are presented in Section~\ref{conclusions}.

\begin{table*}[!htb]
\centering
\begin{scriptsize}
\begin{tabular}{|c|c|c|c|c|c|c|c|c|}
  \hline
  & \cite{OPENiD23} & \cite{Suo13} & \cite{Hassan15} & \cite{StojmenovicFSec05} & \cite{Kanghyo15} & \cite{Shanhe15} & \cite{Yucianga15} & Our work \\
  \hline
  Features, Synergies & No & No & No & No & No & No & \textit{Partial} & \textit{Yes} \\
  Fog - Threats & No & No & No & No & \textit{Partial} & \textit{Partial} & No & \textit{Yes} \\
  Fog - Security & No & No & No & \textit{Partial} & No & \textit{2015} & No & \textit{Q3 2016} \\
  MEC - Threats & No & No & No & No & No & No & No & \textit{Yes} \\
  MEC - Security & No & No & No & No & No & No & No & \textit{Q3 2016} \\
  MCC - Threats & \textit{Partial} & \textit{Partial} & \textit{Partial} & No & No & No & No & \textit{Yes} \\
  MCC - Security & \textit{2013} & \textit{2013} & \textit{2015} & No & No & No & No & \textit{Q3 2016} \\
  \hline
\end{tabular}
\end{scriptsize}
\caption{Contribution of available surveys on Edge security}\label{T:Related}
\end{table*}

\paragraph{Related Work} In recent years, various authors have surveyed and reviewed the state of the art of the security of various edge paradigms, such as mobile cloud computing~\cite{OPENiD23,Suo13,Hassan15} and fog computing~\cite{StojmenovicFSec05,Kanghyo15,Shanhe15}. Such works look to provide a preliminary analysis of the threats that affect the integrity of these paradigms, alongside an overview of the security mechanisms by which to protect all actors and infrastructures. Other works focused on specific areas, such as network security~\cite{Stojmenovic14} and forensics~\cite{WangF15} in fog computing. Moreover, certain authors~\cite{Yucianga15} have also provided an brief overview of the basic features of all edge paradigms. However, as shown in table~\ref{T:Related}, this is the first study to provide a detailed and up-to-date analysis of several subjects from a holistic perspective, including i) the common features, differences, and synergies of edge paradigms, ii) a detailed analysis of the various threat models that target the integrity of all edge paradigms, and iii) a thorough analysis of the state of the art of security in all edge paradigms, including potential synergies among security mechanisms.


\section{Overview of Edge Paradigms}
\label{overview}

\subsection{Fog Computing}
\label{overview:fog}

The concept of Fog Computing was introduced by Cisco Systems in 2012, and in its initial definition it was considered as an ``extension of the cloud computing paradigm (that) provides computation, storage, and networking services between end devices and traditional cloud servers'' \cite{Bonomi12}. Therefore, fog computing does not cannibalize cloud computing, but complements it: the fog architecture facilitates the creation of a hierarchical infrastructure, where the analysis of local information is performed at the `ground', and the coordination and global analytics are performed at the `cloud'. Here, cloud services are deployed mostly at the edge of the network, but they can also be deployed in other locations, such as IP/multiprotocol label switching (MPLS) backbones. In fact, the fog network infrastructure is heterogeneous, where high-speed links and wireless access technologies will coexist~\cite{Bonomi14}.

The initial definition of fog computing was later expanded and revised by various researchers (cf.~\cite{Vaquero14,Yi15}). Although this extended definition is debatable, it reveals all the advances that the fog might introduce. Under this new definition, fog computing does not become a mere extension of cloud computing, but a paradigm of its own. The elements that implement the cloud services, the fog nodes, can now range from resource-poor devices (e.g. end devices, local servers) to more powerful cloud servers (e.g. Internet routers, 5G base stations). Also, all these elements can also be able to interact and cooperate with each other in a distributed fashion. This generates a three-tier architecture (Clients $\Leftrightarrow$ fog nodes $\Leftrightarrow$ Central Servers) where centralized cloud servers coexist with fog nodes but are not essential for the execution of fog services~\cite{Luan15}. Moreover, fog computing also provides support for the creation of federated infrastructures, where multiple organizations with their own fog deployments can cooperate with each other.

Originally, fog computing was defined as a platform that enabled the creation of new applications and services in the context of the Internet of Things (IoT). Examples of such services include hierarchical Big Data analytics systems and smart infrastructure management systems (e.g. wind farms, traffic lights)~\cite{Bonomi12,Vaquero14}. Yet, at present, there are several studies that examined how this paradigm could be used to implement other types of services: low-latency augmented interfaces for constrained (mobile) devices (e.g. brain-computer interfaces using wireless electroencephalogram headsets~\cite{Zao14}, augmented reality and real-time video analytics~\cite{Ha14}), cyber-physical systems~\cite{StojmenovicCG14}, novel content delivery and caching approaches under the context of fog computing~\cite{Jingtao15}, and various vehicle-to-vehicle (V2V) and vehicle-to-infrastructure (V2I) services such as shared parking systems~\cite{Kim15}.

As of 2016, the efforts of creating a set of standardized open fog computing frameworks and architectures have started (cf. Open Fog Consortium \cite{OpenFog15}). These efforts do not need to start from zero, as various researchers have already analyzed what a fog computing architecture could look like. One example is the architecture defined by Sang Chin et al.~\cite{Chin15}. This context-aware infrastructure supports a diversity of edge technologies (e.g. Wi-Fi, LTE, ZigBee, Bluetooth Smart), and also supports network virtualization and traffic engineering through Network Function Virtualization (NFV) and Software Defined Networking (SDN) mechanisms. Other researchers have studied how fog computing could be integrated with existing IoT frameworks, such as OpenM2M \cite{Datta15}. In this particular example, fog nodes are deployed at edge devices such as road side units in vehicular networks, and implement various machine-to-machine services such as lightweight M2M device management systems and M2M sensor measurement frameworks.

There are also various researchers that have already identified not only potential challenges, but also forward-thinking deployments that make use of the fog computing paradigm in novel ways. One example is the need to provide an set of APIs that will allow Virtual Machines (VM) to access to services provided by fog nodes. Using these APIs, VMs can access to local information such as network statistics, sensor data, etc~\cite{Zhanikeev15}. Another example is the deployment of \textit{Airborne Fog Computing} systems -- where flying devices such as drones act as fog nodes and collaborate with each other and with other servers in order to provide various services to mobile users~\cite{Loke15}.


\subsection{Mobile Edge Computing}
\label{overview:mec}

The term \textit{Mobile Edge Computing} (MEC) was firstly used to describe the execution of services at the edge of the network in 2013, when IBM and Nokia Siemens Network introduced a platform that could run applications within a mobile base station~\cite{MEDfirst}. This initial concept only had a local scope, and didn't consider other aspects such as application migration, interoperability, and others. MEC acquired its current meaning afterwards, in 2014, when the ETSI launched the Industry Specification Group (ISG) for Mobile-Edge Computing~\cite{MEDwhite14}. Under this specification, MEC aims to ``provide an IT service environment and cloud-computing capabilities at the edge of the mobile network''. This group also pursues the creation of an open ecosystem, where service providers can deploy their applications across multi-vendor MEC platforms. Once the standard is finished, telecommunication companies will be in charge of deploying this service environment in their infrastructure.

The benefits of deploying cloud services at the edge of mobile networks like 5G include low latency, high bandwidth, and access to radio network information and location awareness. Thanks to this, it will be possible to optimize existing mobile infrastructure services, or even implement novel ones. An example is the Mobile Edge Scheduler \cite{Fajardo15}, which minimizes the mean delay of general traffic flows in the LTE downlink. Moreover, the deployment of services will not be limited to mobile network operators, but it will also be opened to 3\textsuperscript{rd} party service providers as well. Some of the expected applications include augmented reality, intelligent video acceleration, connected cars, and Internet of Things gateways, amongst others \cite{MEDwhite15}.

In order to implement the MEC environment, it is necessary to deploy virtualization servers (i.e. MEC servers) at multiple locations at the edge of the mobile network. Some deployment locations considered by the MEC ISG are LTE/5G base stations (eNodeB), 3G Radio Network Controllers (RNC), or multi-Radio Access Technology (3G/LTE/WLAN) cell aggregation sites -- which can be located indoors or outdoors. Besides, the MEC ISG has suggested that this virtualization infrastructure should host not only MEC services, but also other related services such as Network Function Virtualization (NFV) and Software Defined Networking (SDN)~\cite{MEDwhite15}. Such deployment would reduce the deployment costs, and provide a common management and orchestration infrastructure for all virtualized services.

As of 2016, the ETSI Mobile Edge Computing ISG~\cite{MEDwhite14} has produced a MEC framework and reference architecture, whose functional elements provide support to services such as application execution, radio network information, and location awareness. Besides, there are various studies that are investigating how this service environment could be deployed using both existing and novel technologies. For example, Staring et al.~\cite{Staring13} evaluated three major open source cloud computing platforms (OpenStack, Eucalyptus and OpenNebula), and identified what modules need to be improved in order to deploy the platforms in a mobile network. Puente et al.~\cite{Puente15} analyzed how small cell clouds (clusters of interconnected eNodeB) could be seamlessly integrated in existing LTE-A infrastructures without modifying any existing standards and interfaces. Moreover, Maier and Rimal~\cite{Martin15} studied how fiber optic communication technologies could be used to interconnect all the elements of a MEC environment.


\subsection{Mobile Cloud Computing}
\label{overview:mcc}

Mobile Cloud Computing (MCC) mainly focuses on the notion of `mobile delegation': due to the limited resources available to mobile devices, they should delegate the storage of bulk data and the execution of computationally intensive tasks to remote entities. In the original MCC concept, introduced in 2009, only centralized cloud computing platforms were considered as the most viable solution to implement the remote execution of tasks~\cite{Ali09}. Later, other researchers expanded the scope of MCC. In this new vision, tasks could also be delegated to devices located at the edge of the network~\cite{Bahl12}. At present, both visions of MCC coexist~\cite{Rahimi14}. In this study, we will mostly focus on the latter.

Initially, MCC sought to provide novel solutions to services such as mobile learning, mobile healthcare, searching services, and others~\cite{Dinh13}. Nowadays, many of these services can be implemented in a centralized cloud (e.g. voice-based search) or in the mobile devices themselves (e.g. text-to-speech engines). Nevertheless, the concept of MCC is still relevant, as its potential has not been fully exploited. There are certain applications, such as augmented reality and augmented interface applications, where the existence of an execution platform located at the vicinity of the mobile devices can provide several benefits such as lower latency and access to context information. Moreover, as mobile devices are equipped with functional units such as sensors and high resolution cameras, it is possible to develop novel crowdsourcing and collective sensing applications that make use of the location information~\cite{WangMCC15}.

One of the most active areas of research in the field of MCC is the delegation of tasks to external services~\cite{Rahimi14}. There are various solutions that allow applications to migrate part of their code from the mobile devices to cloud-based computing resources located at the edge. Applications are usually implemented using frameworks like .NET and JVM, which makes the code migration process easier. Some research results allow mobile devices to migrate only part of their code, thus is necessary to statically or dynamically identify the code that needs to be offloaded. Other researchers take a more extreme stance: an entire execution environment (i.e. clone), representative of the mobile device, is created. Then, part of the mobile application (including memory image, CPU state, and others) is loaded into the clone. Finally, some approaches make use of mobile agents infrastructures, where the mobile device create a mobile agent that will acquire/process information on its behalf. There are even approaches, such as the concept of Aqua Computing, that mix the notion of mobile agents and clones~\cite{Magurawalage15}.

Another important research area is the implementation of the cloud-based computing resources located at the edge. There are two major strategies: proximate immobile computing entities (fixed virtualization servers), and proximate mobile computing entities (ad-hoc conglomerate of mobile devices)~\cite{Abolfazli14}. In this article we will focus mostly on the first strategy, but will take into account various aspects of the second strategy.

The core element of the first strategy is the cloudlet. This concept, which was firstly defined by Satyanarayanan et al. in 2009~\cite{Satyanarayanan09}, refers to a small cloud infrastructure located near the mobile users. This small infrastructure can be deployed at business premises (e.g. coffee shops, company buildings), uses persistent caching of data and code instead of hard state, and allows devices to load a small VM overlay over pre-existing full-fledged VM images~\cite{Satyanarayanan15}. There are already proofs-of-concept freely available to the research community~\cite{Elijah15}, including user-centric personal cloudlets. Moreover, several tests have shown that cloudlets improve the response time and the energy consumption of mobile devices (51\% and up to 42\%, respectively~\cite{GaoCloudlet15}) in comparison to centralized clouds.

There are various instances of the second strategy. They all specify a distributed computing platform on a cluster made of nearly devices, which play the role of servers based on cloud computing principles. The elements of the cluster can be mobile devices (cf. Hyrax~\cite{Marinelli09}, FemtoClouds~\cite{Habak15}), IoT devices and entities (cf. Aura~\cite{Aura15}), or a combination of several types of devices. Due to the limited resources available to the devices that form the distributed cluster, this strategy does not make use of virtualization techniques. Instead, some implementations make use of specific parallel algorithms such as MapReduce, while others take a more general approach and allow various types of computationally intensive tasks. In almost all cases, a controller is in charge of receiving the tasks and discover what devices could optimally execute them.


\subsection{Other Approaches}
\label{overview:other}

\begin{figure*}
\centering
\includegraphics[width=350pt]{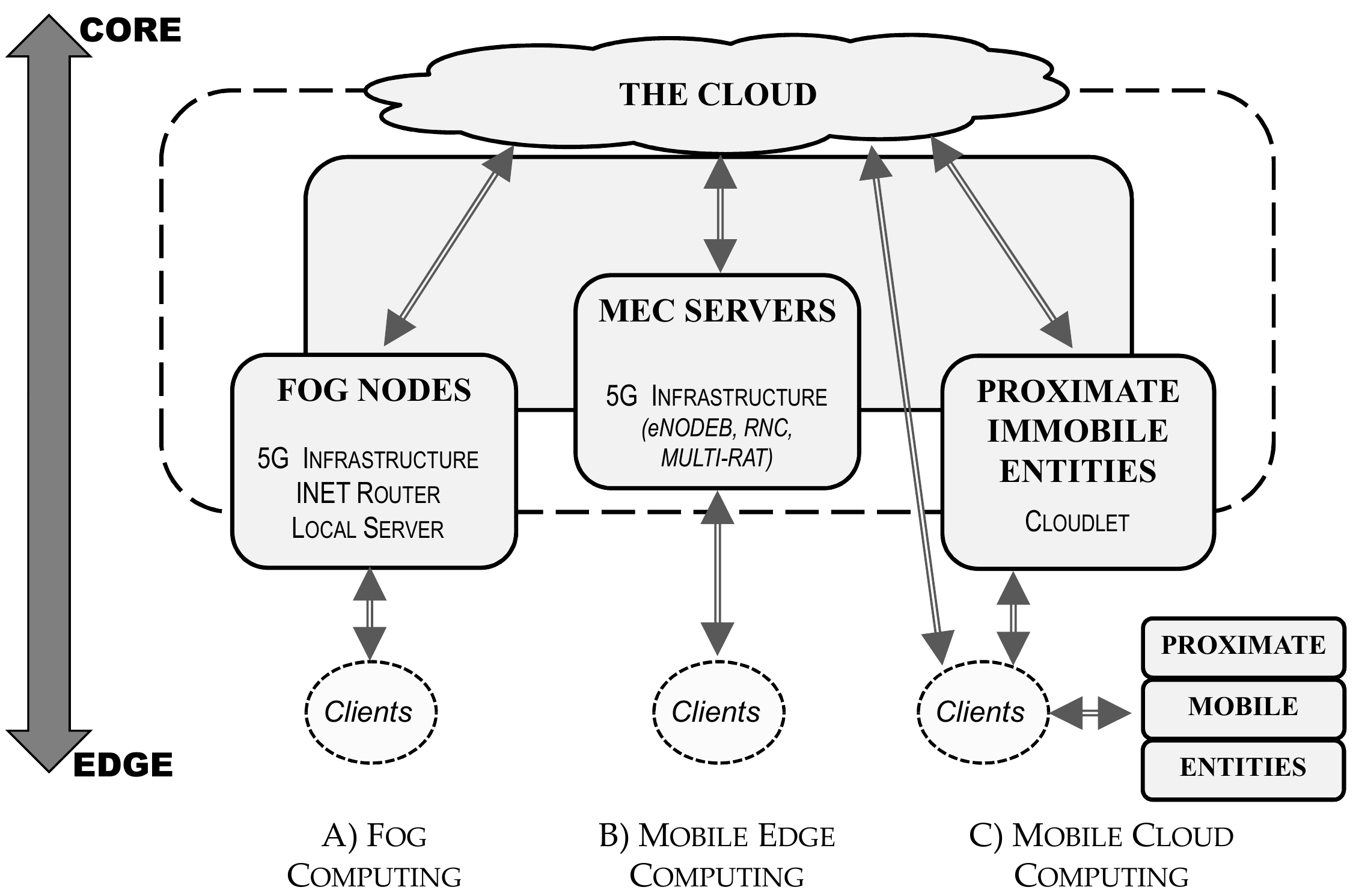}
\caption{Simplified overview of major Edge paradigms}
\label{fig:Comp}
\end{figure*}

As we have seen in the previous sections, there are many paradigms that aim to bring cloud services and resources closer to the users. Although we have provided a summary of the most important ones (cf. Fig.~\ref{fig:Comp}), there are still some minor nascent architectures that are related to these major paradigms. One example is the concept of the \textit{superfluid cloud}, defined by Manco et al. \cite{Manco2015}. In the vision of the superfluid cloud architecture, a set of virtualization platforms with heterogeneous capabilities (from microservers such as Raspberry Pis to larger x86 deployments) are deployed at various points of the network: at the access network level (e.g. 5G eNodeB), at the aggregation network level (e.g. local data centers, network routers), and at the core infrastructure level (e.g. cloud data centers). Besides the deployment of heterogeneous servers, another major differentiator of this architecture is the concept of massive consolidation: the ability to execute a large number (around 10.000) of minimalistic VMs in a single commodity server. These VMs can be deployed and migrated to various points of the network very quickly. This massively distributed, hierarchical architecture enables the creation of on-the-fly services, which might behave as mobile agents if necessary.

Another architecture, known as \textit{edge-centric computing}, was defined by Garcia Lopez et al. \cite{GarciaLopez15}. In their vision, a federation of edge-centric distributed services, deployed across data centers and nano data centers, collaborate with each other in a peer-to-peer fashion. Moreover, the cloud can take an auxiliary role, providing stable resources when necessary. This vision enables the creation of human-centered applications, such as the creation of personal spaces at the edge (e.g. personal information with access control and trust mechanisms managed by the users), social spaces at the edge (e.g. crowdsourcing applications based on user-controlled social activities), and public spaces at the edge (e.g. collaborative information flows where multiple actors -- human and city services -- interact).



\section{Analysis of Features and Synergies}
\label{feat_syn}

\subsection{Features: Similarities and Differences}
\label{feat_syn::feat}

\begin{table*}[!htb]
\centering
\begin{scriptsize}
\begin{tabular}{|c|c|c|c|c|}
  \hline
   & \textit{MEC} & \textit{Fog Computing} & \textit{MCC} & \textit{Cloud} \\
  \hline
  \textit{Ownership} & Telco companies & \multicolumn{2}{|c|}{Private entities, Individuals} & Private entities \\
  \hline
  \textit{Deployment} & Network Edge & Near-Edge, Edge & Network Edge, Devices & Network Core \\
  \hline
  \textit{Hardware} & \multicolumn{2}{|c|}{Heterogeneous servers} & Servers, User devices & Servers \\
  \hline
  \textit{Service} & \multicolumn{2}{|c|}{Virtualization} & Virtualization, Others & Virtualization \\
  \hline
  \textit{Net. Architecture} & \multicolumn{3}{|c|}{N-Tier, Decentralized, Distributed} & Centralized \\
  \hline
  \textit{Mobility} & \multicolumn{3}{|c|}{Yes} & N/A \\
  \hline
  \textit{Latency, Jitter} & \multicolumn{3}{|c|}{Low} & Average \\
  \hline
  \textit{Local Awareness} & \multicolumn{3}{|c|}{Yes} & N/A \\
  \hline
  \textit{Availability} & \multicolumn{4}{|c|}{High} \\
  \hline
  \textit{Scalability} & \multicolumn{3}{|c|}{High} & Average \\
  \hline
\end{tabular}
\end{scriptsize}
\caption{Comparison of features of Edge paradigms}\label{T:PropEdge}
\end{table*}

Table~\ref{T:PropEdge} summarizes the main properties of every major edge paradigm. Some of these properties were introduced in the previous section, while other properties have been gathered from existing reports and research documents (cf.~\cite{MEDwhite15,Puente15,OpenFog15,Luan15,Rahimi14,Dinh13} and others). Note that, for the sake of comparison, this table also includes the properties of the existing centralized cloud computing paradigm.

\paragraph{Similarities} When analyzing the properties of the different paradigms, one apparent conclusion is that these paradigms might come from different backgrounds, but they all have the same basic goal: to bring cloud computing-like capabilities to the edge of the network. They all provide support for some type of multi-tenant virtualization infrastructure (e.g. fog node, MEC server, cloudlet), which is easily accessible through various broadband networks (e.g. fiber optic, wireless communications, high-speed mobile networks). These infrastructures can adjust the provisioning of capabilities to the location and needs of their users, accessing nearby computational resources (e.g. neighbour virtualization pools, distributed mobile devices) if needed. Besides, all paradigms take into consideration the need to monitor the use of the different resources, although the entities in charge of this monitoring and the distribution of these entities varies from paradigm to paradigm.

There are various similarities between all paradigms, as well. One clear example is mobility: as most services are provided locally, it is essential to take into consideration the existence of mobile devices. Every paradigm makes use of several strategies to support user mobility: from mobility management entities located at a higher level in the network hierarchy, to mechanisms that provide support for the migration of VMs. Another example is the network architecture. All paradigms can behave as an extension of the cloud, complementing its services, which enables the creation of a hierarchical multi-tiered architecture. On the other hand, the elements of these paradigms can also behave in a decentralized and distributed way; edge data centers can provide services and take decisions autonomously, and also collaborate with each other without completely depending on a central infrastructure. Moreover, all paradigms pursue the creation of federated infrastructures, where multiple edge infrastructures can coexist and exchange information and services.

These paradigms also provide a similar set of benefits, which are derived mostly from the proximity of the edge data centers. For example, whenever users access to the computing capabilities of their surroundings, both the network latency and the packet delay variation (jitter) are low and predictable. Another benefit is the ability to access to local information (e.g. network conditions, physical aspects of the environment, geographical location), which allows all users and services to be aware of their local context. A further benefit is the scalability of the whole ecosystem. There are various reasons for this. First, nodes can be wide-spread and geographically available in large numbers. Second, it is assumed that the nodes located at a certain site will mostly provide services to local devices. Note that it can be possible to make use of neighbouring nodes, or even use nodes situated at more remote geographical locations or at a higher level in the hierarchy, if the situation requires it. Finally, another important benefit is the high availability of the services. There are two reasons for this: i) node redundancy at a local level, and ii) in certain paradigms (e.g. MEC), the edge data centers are actually hosted by the communication infrastructure (e.g. mobile network infrastructure).

\paragraph{Differences} Obviously, even if all these paradigms have the same goal, they will have some underlying differences on how they want to fulfill that goal. For example, MEC limits the deployment of edge computing platforms to mobile network infrastructures such as 5G. On the other hand, fog nodes can also be deployed at other locations, such as user-managed servers, access points, routers, gateways, etc. As for MCC, it has an even more distributed scope, where in some instances the devices themselves can participate in the service provisioning process. This difference on the deployment and management of the edge data centers influences over who can become a service provider. For example, in MEC, only telecommunication operators can become MEC providers, as they own the mobile network infrastructure where the edge data centers are deployed. In contrast, any user (from companies to tech-savvy end users) can deploy their own fog and MCC nodes, effectively becoming part of the service provisioning ecosystem -- or even creating their own private cloud-like environments.

Another difference, related to the previous point, is the deployment of curated applications. As MEC servers are controlled by telecommunication operators and hosted in their infrastructure, it is possible for third-party service providers to work closely with the operators and develop MEC-specific services. Such services can then be extensively tested and possibly integrated in a customised way. This is also true for fog computing, as certain fog nodes can be deployed in ISP infrastructures (e.g. routers and gateways). Finally, some paradigms, such as MCC, provide some specific services that are not considered by other paradigms. For example, MCC provides support for distributed execution mechanisms that are not related to virtualization, such as the execution of the MapReduce parallel algorithm over constrained devices. Another example is the edge-centric computing vision, which focus on personal spaces (e.g. user-controlled personal networks) and peer-to-peer interactions.

\subsection{Challenges and Synergies}
\label{feat_syn::chasyn}

\begin{table*}[!htb]
\centering
\begin{footnotesize}
\begin{tabular}{|c|c|}
  \hline
  \textit{Challenge} & \textit{Description} \\
  \hline
  Infrastructure & Interoperability; Monitoring; Accountability \\
  \hline
  Virtualization & VM lifecycle; Container and context awareness \\
  \hline
  Resources \& Tasks & Resource location; Task scheduling; Offloading \\
  \hline
  Distribution & Cooperation; N-tier management; `Soft state' \\
  \hline
  Mobility & Connectivity; Seamless handoff \\
  \hline
  Programmability & Usability; Session management \\
  \hline
\end{tabular}
\end{footnotesize}
\caption{Common challenges in Edge paradigms}\label{T:ChEdge}
\end{table*}

Due to the similarities between all edge paradigms, they have several major challenges in common, which are summarized in Table~\ref{T:ChEdge} (cf.~\cite{Yi15, WangMCC15, MEDwhite15} and others). The common denominator of most of these challenges is the decentralized and distributed nature of these paradigms, in contrast with the centralized nature of the cloud computing paradigm. The decentralization and proximity of the service infrastructure to the edge brings various benefits (e.g. low latency, scalability), but it also brings new issues that must be carefully considered. Examples of such issues are the mobility of the various entities (including the service infrastructure itself~\cite{Loke15}) and the need to synchronize `soft' and `hard' states within a multi-tiered architecture. The distributed nature of the service infrastructure, where several edge data centers that might be owned by different infrastructure providers should be able to collaborate with each other, imposes other challenges as well. It is necessary to develop standards that specify how the different elements of the architecture can collaborate with each other, and also how the VMs can access certain information (e.g. context and host information) regardless of their deployment place. Precisely, regarding virtualization, it is essential to provide support for an optimized VM lifecycle -- where their creation, deployment, and migration is as lightweight as possible. Moreover, as resources are distributed over various entities and locations, there must exist a set of mechanisms that enable the discovery and orchestration of such resources, including their monitorization.

While all paradigms share these common challenges, it still is necessary to consider the nuances of every paradigm (e.g. the features of their underlying protocols, their specific use cases) when researching and developing novel solutions. In fact, due to technical, economic and political reasons, it is clear that multiple standards and specifications will coexist, with their own solutions to existing problems. Yet, even if there are several structural differences between the paradigms, it does not mean that they should exist in a vacuum, ignoring the advances in other related fields. Due to the similarities between the paradigms (cf. Section~\ref{feat_syn::feat}), it is safe to assume that there will be mechanisms and platforms that can provide a generic solution to a shared problem. Such solutions can then be adapted to other edge paradigms. In fact, this assumption is supported by the existing state of the art, where there are already various mechanisms that have been developed with a certain paradigm in mind, but that can be applied to other paradigms as well. We will provide various instances of such synergies in the following paragraphs.

One area where clear synergies between paradigms can be found is the management of VMs. There are already various works, in the areas of fog computing~\cite{Kavvadia15} and MEC~\cite{Oueis15}, that define and solve optimization problems whose goal is to improve the distribution of VMs over a set of local clusters, minimizing the utilization of resources. These mechanisms also define when VMs need to replicate, migrate and be merged. As the underlying assumptions made by these algorithms mainly demand a decentralized infrastructure able to access local information, they can be adapted to any edge paradigm. Another important aspect is the cost of setting up and deploying new VMs. Precisely, the advances in areas such as the superfluid cloud~\cite{Manco2015}, where thousands of minimalistic VMs can be deployed in a commodity server with minimal latency, can be adapted and used in other paradigms as well.

Another area of research whose solutions might be applied to other paradigms is resource offloading, which is one of the most studied areas in the field of MCC. As mentioned in Section~\ref{overview:mcc}, there are various strategies that allow user devices to delegate their tasks to external servers (remote procedure call frameworks, code migration, clone deployment, mobile agents). Yet there are situations where not an user device, but a VM might want to delegate part of its tasks. Instances of such situations are user mobility (e.g. specific tasks -- not the whole VM -- are migrated to the closest node to the user), task optimization (e.g. network intensive tasks are kept at a local level, and computationally intensive tasks are sent to a more capable cloud system~\cite{Ottenwalder13MigCEP}), user empowerment (e.g. the creation of user spaces in the network through clones in aqua computing~\cite{Magurawalage15}), and others. Many delegation strategies only require the existence of a virtualization platform where tasks can be delegated~\cite{WangMCC15}, thus a lot of research on this particular area can be adapted to satisfy the needs of applications deployed in other edge paradigms.

Other examples include user mobility, context awareness and location of resources. There are various algorithms, such as~\cite{Hong2013,Ottenwalder2014}, that try to predict the location of (potential) users in order to deploy expected resources in advance. The core algorithms that implement the migration plan only require a distributed architecture of computing platforms that are able to communicate with each other with as less latency as possible. As for context awareness, several works in this area analyze how local hardware awareness can help VMs to understand the limits of their own containers~\cite{Zhanikeev15}, or how VMs can make use of hardware acceleration technologies such as graphic computing units~\cite{Paglierani15}. These works are essential in a context where virtualization servers are heterogeneous, and VMs might need to dynamically adjust their behaviour. About the location of resources, there have been various works in the area of pervasive computing (cf.~\cite{Romer10}) that could be applied to fog, MEC, and other paradigms. Many of these search engines make use of a N-tiered hierarchy, where the lowest layer of the hierarchy stores information about the local contexts. Algorithms such as Bloom filters are then used to represent a set of keywords in order to reduce the communication overhead.

There are also several application scenarios that have been defined for only one paradigm, but due to their requirements (e.g. support for decentralized and distributed execution platforms) they could be implemented in other related paradigms as well. Scenarios that have been defined mostly for fog computing, such as IoT node pairing services~\cite{Abedin15}, context-aware data analytics platforms~\cite{Manolopoulos14}, and emergency notification mechanisms~\cite{Aazam15} could also be deployed in MEC and cloudlet infrastructures. Lastly, there are various supporting services that, even if they have been developed with a single paradigm on mind, they can be adapted to other paradigms. The network store concept, developed by Nikaein et al.~\cite{Nikaein15} is one of such supporting services. This research work introduces a digital distribution platform for MEC, which provides Virtualized Network Functions (VNFs), or slices, that enable 5G application use-cases. Although these slices cannot be directly deployed in other related paradigms due to the differences in the underlying protocols, the concept of a digital distribution platform and its architectural elements (service and business layer, slice orchestrator and manager layer) can be adapted and deployed in other virtualization platforms. Precisely, due to the distributed and collaborative nature of edge paradigms, the implementation of a network store can serve as a catalyst for the rapid deployment of edge applications, as it can behave as a repository of functionality and knowledge.

As a final note, we want to emphasize that, even if every paradigm pursues the creation of their own standards and their own service infrastructure, this doesn't mean that it is not possible for them to collaborate with each other -- or even being integrated with each other. For example, cloudlets are traditionally associated to MCC, but they can become a technology enabler for both fog computing and MEC. Also, as both fog computing and MEC aim to provide support for federated services and interactions with different providers (e.g. through a set of open APIs), it can be possible to create applications that make use of both edge paradigms, or even deploy middleware platforms that will connect various edge paradigms at an infrastructure level.


\section{Security Threats}
\label{sec_thre}

There are several challenges that must be overcome in order to create an ecosystem where all actors (end users, service providers, infrastructure providers) benefit from the services provided by edge paradigms. Not surprisingly, one of the greatest challenges is security. In this section, we will a) review why security is a very important factor in this particular context (section~\ref{sec_thre::whysec}), b) analyze the specific threats that can target edge paradigms (section~\ref{sec_thre::land}), and c) introduce the requirements and challenges of the security mechanisms that should be applied to this particular context (section~\ref{sec_thre::mech}).

\subsection{The Importance of Security in Edge Paradigms}
\label{sec_thre::whysec}

As aforementioned, one of the greatest challenges for the creation of a edge paradigm ecosystem is security. There are several reasons for this. First, at the core of most edge paradigms, there are several enabling technologies such as wireless networks, distributed and peer-to-peer systems, and virtualization platforms~\cite{Vaquero14}. It is then necessary not only to protect all these building blocks, but also to orchestrate the diverse security mechanisms. This is by itself a complex issue, as we need to create an unified and transversal view of all the security mechanisms that allows their integration and interoperability.

Second, the whole is greater than the sum of its parts: by assuring the security of all the enabling technologies, we do not assure the security of the whole system. Once cloud computing-like capabilities are brought to the edge of the network, novel situations arise (e.g. collaboration between heterogeneous edge data centers, migrating services at a local and global scale) whose security has not been widely studied. Besides, we also need to consider the specific requirements of this particular context (cf. Section~\ref{feat_syn::feat}), which might affect the kind of security mechanisms that could be deployed. For example, the security mechanisms should be as autonomous as possible and not depend on the continuous existence of a centralized infrastructure. There are two main reasons for this: not only there will be situations (e.g. malicious attacks, intermittent connectivity, distributed applications) where no centralized control system is available, but it is also necessary to take into account the latency of the security mechanisms. Another example is the technological limitations of the elements of the infrastructure. For example, certain edge data centers might be composed of microservers (e.g. Raspberry Pi) that lack the hardware protection mechanisms of commodity servers~\cite{Manco2015}, or include legacy edge devices with limited connectivity -- which restricts the authentication protocols that can be deployed~\cite{StojmenovicFSec05}. Moreover, the security mechanisms need to consider the existence of mobile devices, which can make use of the edge data centers anytime and anywhere.

Third, we need to keep in mind that, in addition to the security threats that will appear due to the specific features of edge paradigms, the whole system also inherits the security threats that are present in their building blocks and in the application scenarios. And this is no trivial issue, because these threats are, in fact, very significant. A clear example of this is the Internet of Things, which is the main \textit{raison d'etre} of fog computing~\cite{Bonomi12} -- and a major use case in all edge paradigms. It is also considered as a combination of ``the worst-of-all-worlds'' in terms of security: not only we need to combine and protect multiple layers of technologies (from network to mobile to cloud~\cite{Miessler15}), but also provide global connectivity and accessibility in a heterogeneous ecosystem~\cite{Roman2013}. This situation generates a considerable attack surface, that in turn also affects all paradigms that make use of the IoT.

Finally, the impact than a successful attack might cause in our society is quite considerable. The number of application scenarios where edge paradigms can be applied is huge. In fact, almost any aspect of our daily lives can be influenced by applications deployed in these infrastructures: Our private information (e.g. photos, medical reports), our daily routines (e.g. transportation, shopping), our enterprise ecosystems (e.g. industries, supply chains), our critical infrastructures (e.g. energy, emergency systems), etc. Without proper security and privacy mechanisms, the benefits of edge paradigms will be quickly overshadowed by the damage caused by malicious adversaries.

\subsection{Threat landscape}
\label{sec_thre::land}

Once we have understood the importance of security in the context of edge paradigms, it is time to analyze what are the specific threats that can target these paradigms, and the extend of the damage they can cause. In the near future, this analysis will help us in the development of security mechanisms that can adequately protect the whole ecosystem against such threats. Besides, it will also allow us to understand what are the particularities of every edge paradigm, as they have subtle differences that will affect the implementation and deployment of the security mechanisms.

However, before analyzing the threats, it is necessary to examine how the lack of a global perimeter affects the security of edge paradigms. As we have seen in previous sections, even in the most closed paradigm -- mobile edge computing -- the whole ecosystem will not be controlled by one single owner. Even more, edge data centers are capable of providing services without continuously depending on a central infrastructure. Therefore, all relevant assets, including the network infrastructure, the service infrastructure (e.g. edge data centers, core infrastructure), the virtualization infrastructure, and the user devices, are controlled not by a single entity but by various actors (including, in some cases, end users) who need to cooperate with each other. A consequence of this situation is that every element of the infrastructure can be targeted or subverted at any moment. In fact, this ``anything, anytime'' principle is also inherited from some of the underlying building blocks and application scenarios, such as the Internet of Things~\cite{Roman2013}.

Having said that, the ``anywhere'' principle (attacks can be performed from anywhere) does not fully apply to this particular context. The cause of this is the geographical location of the edge data centers. One of the basic tenets of these paradigms is that cloud computing capabilities are basically provided in close proximity to end users. As a result, a edge data center (e.g. fog nodes, MEC servers) will provide services mostly to local entities (e.g. mobile users located at the vicinity, entities inside a building). There are a few exceptions to this rule, such as virtual machines that act like agents and migrate to other infrastructures away from their physical counterpart, or specific local services that are requested by remote entities. This particularity of edge paradigms is a double-edged sword: on the one hand, it limits the impact of an attack to the local environment. On the other hand, if one adversary can control one edge data center, it might be able to control almost all the services that are provided in that geographical location.

There is another consequence of the lack of a global perimeter: the nature of the different attacker profiles that will target edge paradigms. Even if traditional `external' attackers will exist (i.e. adversaries that do not control any element of the whole infrastructure), there will exist many adversaries that will control one or more elements of the infrastructure: user devices, virtual machines, servers, sections of the network, even whole edge data centers. This situation is similar to the current Internet, where malicious adversaries can take control of existing elements or deploy their own. These adversaries are both `internal' and `external', as they control one part of the infrastructure but not the others. Note that these attackers can still try to influence other healthy sections of the infrastructure. Examples are the injection of bogus information during a collaboration process, or the deployment of malicious virtual machines that, like viruses, will try to exploit vulnerabilities in their hosts. Needless to say, traditional `internal' attackers (i.e. undercover adversaries such as disgruntled employees that are officially allowed to access certain elements of part of an infrastructure) will also exist within this ecosystem.

\subsubsection{Threat Model}
\label{sec_thre::land::model}

\begin{table*}[!htb]
\centering
\begin{footnotesize}
\begin{tabular}{|c|c|}
  \hline
  \textit{Asset} & \textit{Threats} \\
  \hline
  Network infrastructure & Denial of service, man-in-the-middle, rogue gateway \\
  \hline
  Edge data center & \makecell{Physical damage, privacy leakage, privilege escalation, \\ service manipulation, rogue data center} \\
  \hline
  Core infrastructures & Privacy leakage, service manipulation, rogue infrastructure \\
  \hline
  \makecell{Virtualization \\ infrastructure} & \makecell{Denial of service, misuse of resources, privacy leakage, \\ privilege escalation, VM manipulation} \\
  \hline
  User devices & Injection of information, service manipulation\\
  \hline
\end{tabular}
\end{footnotesize}
\caption{Categorization of threats in Edge paradigms}\label{T:Threats}
\end{table*}

After reviewing the nature and scope of the potential attackers, we can finally provide an analysis of the threats. For this analysis, we will enumerate the most important assets of edge paradigms, and then summarize the attacks that can be launched against such assets. Note that some of the threats that affect edge paradigms will be the same threats that affect traditional data centers, as both of them share various assets (e.g. server farms, networking infrastructure). Still, in our analyses we need to consider the specific decentralized and distributed nature of edge paradigms, plus the existence of additional services such as interoperability and mobility support, location awareness, and others. Therefore, not only the impact of certain common threats will be different (e.g. an attack to an edge data center will mostly impact the services related to that geographical area), but also novel threats will arise.

A summary of this threat classification can be found in table~\ref{T:Threats}. Note that this classification will be defined in a way that it can be applied to all edge paradigms -- we will explicitly explain the particularities of every major paradigm afterwards, in Section~\ref{sec_thre::land::particularities}.

\paragraph{Network Infrastructure} As aforementioned, edge paradigms make use of various communication networks to interconnect their elements: from wireless networks to mobile core networks and the Internet. An adversary can try to target any of these communication infrastructures.
    \begin{itemize}
      \item \textit{Denial of Service (DoS).} All communication networks are vulnerable to several DoS attacks, such as distributed denial-of-service (DDoS) attacks and wireless jamming. Yet the scope of these attacks is limited. Attacks against the edge networks will only disrupt the vicinity of the affected networks. Also, an attack to the core infrastructure might not completely disrupt the functionality of the edge data centers, as their protocols and services can be designed to work in an autonomous or semi-autonomous way.
      \item \textit{Man in the the Middle.} Malicious adversaries can be able to take control of a section of the network, and then launch attacks such as eavesdropping and/or traffic injection. The practicality of this particular threat was demonstrated by Stojmenovic et al.~\cite{StojmenovicFSec05}. In this particular case, a gateway that interconnected two 3G and WLAN networks was compromised, and the adversary gained access to the network interfaces. This attack is not only very stealthy but also very dangerous, as it can affect all the elements (e.g. information, virtual machines) that traverse that particular node.
      \item \textit{Rogue Gateway.} The open nature of several edge paradigms, where even user-owned devices can become full-fledged participants (e.g. personal cloudlets, mobile devices participating in a cluster of nearby devices), create a scenario where malicious adversaries can deploy their own gateway devices. This particular threat produces the same outcome as the Man-in-the-Middle attack (e.g. the ability to eavesdrop and/or inject traffic), even if the means are different (compromising versus deploying).
    \end{itemize}
\paragraph{Service Infrastructure: Edge data center} The edge data center hosts the virtualization servers and several management services, amongst others. However, for an external adversary, the attack surface of a edge data center is quite considerable: from multiple public APIs that provide services to all actors (e.g. users, virtual machines, other data centers) to other access points such as web applications. Note that the specific threats related to the virtualization infrastructure will be described later.
    \begin{itemize}
      \item \textit{Physical damage.} In certain paradigms, the elements of the service infrastructure might not be guarded or protected against physical damage. Clear examples are fog nodes managed by small businesses and user devices forming clusters. For this particular threat, it is necessary for the attacker to be in the vicinity of the device in order to destroy it. As a result, there is a very high probability that this kind of attack will be witnessed by various observers. Moreover, the impact of this particular attack is limited to a local scope: only the services associated to a particular geographical location will be disabled.
      \item \textit{Privacy leakage.} Both internal adversaries and honest but curious adversaries can try to access the flow of information that traverse the edge data center. Nevertheless, the scope of these attacks is limited: An edge data center mainly stores and processes information from the entities that are located at its vicinity, although in some cases (e.g. distributed storage services, migrating virtual machines) it can deal with data coming from other locations. Note, however, that these edge data centers might be able to extract more sensitive information about a user thanks to their awareness of the context~\cite{Shanhe15}.
      \item \textit{Privilege escalation.} The considerable attack surface of these edge data centers allows external adversaries to try to take control of various of its services. This is facilitated by the fact that edge data centers can be managed by professionals with limited security training, or even hobbyists. These infrastructures might be misconfigured, or even lack proper maintenance. Note that this attack can also be performed by internal adversaries that abuse of their privileges and take advantage of their insider knowledge.
      \item \textit{Service manipulation.} Once an adversary has gained control of certain sections of the edge data center, either by privilege escalation or by abusing his own privilege as a legitimate administrator, it can manipulate the services of the data center. As a result, the adversary can launch several types of attacks, such as selective denial of service attacks and selective information tampering, amongst others.
      \item \textit{Rogue data center.} In this threat, an adversary is able to control an entire edge data center through various means, such as privilege escalation or deploying his own malicious infrastructure. This creates a very dangerous scenario, as the adversary i) has complete control of all the services that are provided in a geographical location, ii) has access to all information flows that are directed to the rogue data center, and iii) can manipulate all interactions with external systems (e.g. migrating virtual machines, service requests from remote entities).
    \end{itemize}
\paragraph{Service Infrastructure: Core infrastructures} All edge paradigms can be supported by several core infrastructures, such as mobile core management systems and centralized cloud services. It is then necessary to analyze what are the specific threats that target these upper layers in this particular context. It should be noted that, in certain paradigms (e.g. MEC), the core infrastructure will be managed by the same companies (e.g. mobile network operators) that deploy the edge data centers. Besides, we should not assume that all interactions with a cloud provider can be completely trusted, due to cyber-crimes~\cite{Raymond10} and other reasons (e.g. government intrusion~\cite{Landau14}). A complete taxonomy of general threats that target cloud providers is available elsewhere~\cite{Viktoria15}.
    \begin{itemize}
      \item \textit{Privacy leakage.} There are no guarantees that all information flows that are processed and stored in the upper layers of our edge infrastructures will not be accessed by unauthorized entities or honest but curious adversaries. Note, however, that these internal adversaries might not have access to the whole information set, including raw measurements. The reason is simple: as the lower layers, the edge data centers, will process the local information, it is probable that the upper layers will only receive a subset of said information. In addition, edge paradigms allow edge data centers to exchange information directly with each other, bypassing the central systems.
      \item \textit{Service manipulation.} An internal adversary with enough privileges can try not only to manipulate the information flow, but also to instantiate rogue services that will provide bogus information (e.g. fake management information, historic data) to other partners. But this particular threat follows the same principle as the privacy leakage threat: these internal adversaries will not be able to influence the whole ecosystem, due to the decentralized and distributed nature of edge paradigms.
      \item \textit{Rogue infrastructure.} This threat assumes that certain elements of the core infrastructure can be targeted by specialized adversaries. Such attacks will be able to take control of some services of the upper layers of the infrastructure, causing havock on the whole ecosystem. Although the chances of an adversary successfully launching this attack are extremely low, it is still necessary to have this scenario in mind for especially sensitive situations, where specialized security and fault tolerance mechanisms need to be deployed.
    \end{itemize}
\paragraph{Virtualization Infrastructure} Within the core of all edge data centers, we can find a virtualization infrastructure, which enables the deployment of cloud services at the network edge. Like all other assets, this infrastructure can be exploited in several ways. Besides, we also need to consider that the virtual machines themselves might be controlled by malicious adversaries who are trying to misuse or exploit the resources available to them.
    \begin{itemize}
      \item \textit{Denial of Service (DoS).} A malicious virtual machine can try to deplete the resources (including computational, network and storage resources) of the host where it is running. This threat is quite significant for this particular context, as most edge data centers will not have the resources that are available to other cloud infrastructures.
      \item \textit{Misuse of resources.} A malicious virtual machine can execute various malicious programs that do not target the edge data center where it is hosted, but other local or remote entities. For example, a malicious virtual machine can search for vulnerable IoT devices in the local environment. It can also execute programs for cracking passwords, or host botnet servers.
      \item \textit{Privacy leakage.} Due to requirements in their design, most virtualization infrastructures located at edge data centers are not completely transparent: they can actually implement various APIs that provide information about the physical and logical environment, such as the state of the local network. However, if these APIs are not protected, a malicious virtual machine can be able to obtain sensitive information about the execution environment and the surroundings of the edge data center.
      \item \textit{Privilege escalation.} Malicious virtual machines can also try to take advantage of vulnerabilities in their hosts. There are various outcomes of this attack: from isolation failures, where the malicious VM succeeds at manipulating other VMs, to escalation of privileges, where the malicious VM takes control of certain elements of the host. This problem is exacerbated by the fact that virtual machines can migrate from one data center to the other due to various reasons (e.g. users moving from one location to the other, virtual machines acting as agents).
      \item \textit{VM manipulation.} A host system that is being controlled by an adversary (e.g. a malicious insider with enough privileges, a VM that has escalated privileges), can launch several attacks to the VMs that are running inside it. These attacks can range from the extraction of information to the manipulation of the computational tasks are being executed within the VM. Moreover, the adversary can also infect the VM with logic bombs, malware or other malicious elements that will compromise the security of other data centers once the VM migrates to other physical locations.
    \end{itemize}
\paragraph{User devices} The devices controlled by the users are also important elements of the whole ecosystem. They not only consume services, but also can become active participants that provide data and participate in the distributed infrastructure at various levels. However, there will be also rogue users that might try to disrupt the services in one way or another. Note, however, that the scope of these threats is quite limited: in this context, users can only influence their immediate surroundings.
    \begin{itemize}
      \item \textit{Injection of information.} Any device that is controlled by an adversary can be reprogrammed to distribute fake information when queried (e.g. vehicles reporting wrong values, users providing fake data to crowdsourcing services). Note that a device might also provide bogus values due to an anomaly in their sensors or internal systems.
      \item \textit{Service manipulation.} There are some cases where a device might participate in the provisioning of services. For example, a cluster of devices controlled by a virtual machine located at an edge data center can act as a distributed computing platform. Yet if an adversary gains control of one of these devices, it can be able to manipulate the outcome of the service.
    \end{itemize}

\subsubsection{Differences Between Paradigms}
\label{sec_thre::land::particularities}

In this section, we will make use of the features defined in Section~\ref{feat_syn} to analyze how the threats presented in the previous section affect all paradigms. One feature that has a noticeable effect on the impact of the previous threats is the \textbf{ownership} of the infrastructure. In some paradigms, such as \textit{mobile edge computing}, one single company (the mobile network operator) controls not only various edge data centers located at different geographical locations, but also part of the core networks that are connected to those data centers (i.e. the mobile network infrastructure). In principle, this infrastructure is well maintained, with a consistent security policy and guarded against physical and virtual intruders. Thanks to this, the attack surface should be smaller, which decreases the chances of an adversary destroying or gaining control of part of the service infrastructure. Other paradigms, such as \textit{fog computing} and \textit{mobile cloud computing}, allow small companies (e.g. stores) to deploy their own edge data centers, or even allow users to become active participants in the provisioning of services. This creates a more heterogeneous ecosystem, which will probably be less protected than the infrastructures deployed by big companies due to various reasons (e.g. deficient maintenance, limited physical protection).

However, having a large segment of the service infrastructure managed by one single company has its drawbacks, too. One clear example is the impact of a successful attack. Once an adversary has taken control of a section of the infrastructure, he becomes an internal attacker within that infrastructure. If the necessary contingency mechanisms are not in place, he might try to gain more privileges and/or exploit further vulnerabilities in order to gain even more influence. Moreover, if an insider adversary takes control of certain elements of the core network of that company, it can be able to manipulate large sections of the whole ecosystem. On the other hand, if an adversary takes control of an edge data center managed by a small company or a tech-savvy individual, his reach will be limited to the scope of that particular edge data center.

Another feature that has some influence on the security threats is the \textbf{hardware} used to implement the cloud services in the edge data centers. Paradigms such as \textit{mobile cloud computing} and concepts such as the \textit{Superfluid Cloud} can make use of microservers (Raspberry Pi) and user devices (mobile phones) to provide their services. At present, it is still necessary to analyze how the hardware extensions of certain microcontrollers can be used to guarantee a secure virtualization environment~\cite{Lengyel14}. Regarding the hardware used in paradigms such as \textit{fog computing}, the commodity servers used in small-scale deployments can make use of the same security mechanisms as the commodity servers used in cloud deployments~\cite{Pek2013}. Note, however, that some small-scale deployments might lack experienced staff, and as a result there will be some processes (e.g. definition of security policies, separation of roles, storage of logs in separate physical storage) that might not be properly implemented or maintained.

Regarding the \textbf{deployment} of the elements of the infrastructure, we have already mentioned that certain instances of the \textit{mobile cloud computing} paradigm allows the creation of clusters of devices at the very edge of the network, and that these clusters provide services through mechanisms such as parallelization. Because of this, the MCC paradigm has his own extra set of security challenges~\cite{Suo13}, such as the impact of malware in the user devices, the identification and authentication of the different peers, and the existence of DoS attacks that target honest participants. Besides, we need to mention one aspect that is strongly linked to the \textbf{network architecture}. All paradigms support the creation of a hierarchical multi-tiered architecture, where different elements (user devices, edge data centers, core infrastructures) have different roles. As such, certain security services (e.g. authentication, monitorization) can be deployed in a more centralized or a more distributed way. Every approach has its own advantages and disadvantages. For example, if a centralized service is rendered unavailable or is controlled by an adversary, the whole infrastructure will collapse unless contingency mechanisms are in place. Finally, we also need to consider that certain paradigms will make use of their own \textbf{protocols} and \textbf{services}, such as the Small Cell as a Service (SCaaS) elements in MEC environments, which will have their own security requirements (cf.~\cite{Vassilakis2016}).

\subsection{Security Mechanisms}
\label{sec_thre::mech}

In order to create an effective layer of defense against the different threats, it is crucial to deploy various types of security services and mechanisms. In this section, we will introduce the security services and mechanisms that should be integrated in all edge paradigms, alongside with a brief overview of their requirements and challenges in this particular context. Note that all security mechanisms need to take into account various common requirements and constraints, such as reducing the latency of their operations as much as possible, supporting mobile devices and other mobile entities (e.g. virtual machines), achieving technical, functional, and semantic interoperability, managing the limitations of existing technologies, and providing support for disconnected operations.

\paragraph{Identity and Authentication} In all edge paradigms, there are multiple actors (end users, service providers, infrastructure providers), services (virtual machines, containers), and infrastructures (user devices, edge data centers, core infrastructures) interacting in an ecosystem where multiple trust domains coexist. This situation brings numerous challenges, as not only we need to assign an identity to every entity, but also we need to allow all entities to mutually authenticate each other. Without these security mechanisms, it would be very easy for external adversaries to target the resources of the service infrastructure with impunity. Moreover, internal adversaries would not leave a trail of evidence behind their malicious acts.

In this context, it is necessary to explore identity federation mechanisms and inter-realm authentication systems, which should be interoperable with each other. Besides, due to various requirements (latency, availability of a central server), it is also desirable that an entity can provide a proof of its identity without contacting a central server (e.g. presenting valid and trusted attributes). Note, however, that in some cases parts of the infrastructure can be managed by end-users (e.g. personal cloudlets), and even interact in a peer-to-peer fashion. Therefore, we should study the applicability of distributed authentication mechanisms.

\paragraph{Access Control Systems} The existence of an authorization infrastructure is equally important for edge paradigms, as it is essential to check the credentials of the various entities in order to authorize their requests to perform certain actions (e.g. service providers deploying virtual machines, virtual machines accessing edge data center APIs, edge data centers interacting with each other). If there are no authorization mechanisms in place, anyone without proper credentials can misuse the resources of the virtualization infrastructure. Users would be able to impersonate administrators and control the services of the infrastructure. Malicious attackers would be able to access any resources, including proprietary and/or personal information. The possibilities would be limitless.

Due to the inherent features of edge paradigms, it is crucial to deploy an authorization infrastructure in every trust domain, so as to allow the owners of such domains to disseminate, store and enforce their own security policies. Such infrastructures should be able, in principle, to process the credentials of any entity if there is a trust relationship between them. Moreover, it should be also possible to take into account various factors, such as the geographical location and the resource ownership, in the definition of the authentication policies. For example, migrating virtual machines might be allowed to use additional resources from the virtualization infrastructure if they hold certain privileges (e.g. owned by local law enforcement agencies).

\paragraph{Protocol and Network Security} If the network infrastructure is not protected, the whole service ecosystem will be threatened by internal and external malicious adversaries. It is then necessary to protect the myriad of communication technologies and protocols that are used by edge paradigms. For example, there are various wireless communication technologies (e.g. Wi-Fi, 802.15.4, 5G, Sigfox, LoRa) that might be used to serve local customers. Therefore, edge data centers and their administrators need to understand and make use of the security protocols and extensions implemented by such technologies. Also, edge paradigms need to configure and integrate the security protocols that are used by the core infrastructures (e.g. public Internet, mobile network infrastructure). Moreover, we need to provide network isolation among tenants in the virtualization infrastructure, among other protection mechanisms.

Here, there are various challenges that need to be addressed. For starters, it is necessary to adequately configure the different elements of the network infrastructure. Yet all these elements will be deployed in different geographical locations, which will be managed by different administrators. Besides, there will situations where entities that belong to different trust domains (e.g. edge data centers from different infrastructure owners) will interact with each other. In this very heterogeneous scenario, we need to establish a secure connection between entities that might even use different communication technologies. There are other aspects that are just as important, such as achieving a dynamic balance between the strength of the security mechanisms and the overall quality of service of the network.

\paragraph{Trust Management} Another security mechanism that is of great importance for edge paradigms is trust. In this context, the concept of trust goes beyond the idea of ``not knowing who I am interacting with'', which is mostly solved by implementing authentication mechanisms and establishing trust relationships between trust domains. The reason is simple: we also have to deal with the concept of uncertainty, or ``not knowing how my partner is going to behave''. All entities have a variety of collaborating peers at their disposal: users can have various service providers available in their vicinity, service providers can choose from many infrastructure providers, and so on. However, such peers might not meet our expectations: the service latency might be high, the anomaly detection rate might be low, or the data might be inaccurate. There are even worse situations: peers might behave egoistically or maliciously.

It is then necessary to seriously consider the deployment of trust management infrastructures in this context. The benefits are numerous: from improving the decision-making processes of all entities (e.g. migrate high priority virtual machines to nearer edge data centers with higher reputation), to enhancing the management of personal data (e.g. reduce the granularity of the information that is transmitted to low reputation entities), amongst others. There are many challenges, though. All trust management infrastructures should be able to exchange compatible trust information with each other, even if located at different trust domains. Another problem lies with the storage and dissemination of trust information, as it should be accessible anywhere, anytime, with as less latency as possible. Moreover, due to the dynamic nature of the infrastructures, non-malicious entities might find themselves with a low reputation due to temporary reasons, thus it is necessary to find a balance between punishment and redemption.

\paragraph{Intrusion Detection Systems} Talking about malicious entities, we have already seen in Section~\ref{sec_thre::land::model} that external and internal adversaries can attack any entity at any time. Without proper intrusion detection and prevention mechanisms, any successful attacks will go undetected, slowly undermining the functionality of the whole infrastructure. It is then necessary to ensure that the whole infrastructure is covered by such defense mechanisms. Fortunately, we also have seen that the ``anywhere'' principle does not completely apply to these paradigms: the impact of most attacks is usually limited to a local environment. Therefore, local infrastructures, such as edge data centers, can be in charge of monitoring all their elements -- network connections, virtual machines, etc -- and their surroundings. Besides, these local infrastructures can also cooperate with each other or with core infrastructures located at a higher level in the network hierarchy. This way, it can be possible to detect attacks that target large sections of the service infrastructure.

However, the challenges of running an interconnected network of detection and prevention mechanisms in a heterogeneous, decentralized and distributed infrastructure are numerous. The specific attacks that can be launched against edge paradigms need to be understood. If a database of attacks is used (e.g. for signature-based IDS), it needs to be updated and protected at all times. A balance between local and global defense mechanisms needs to be achieved, and a global monitoring infrastructure that encompasses multiple layers and/or trust domains needs to be developed. Moreover, all defense mechanisms, regardless of their location, must be able to exchange information with each other in an interoperable format. Such information should be permanently available in order to detect more persistent threats. Finally, the defense mechanisms must behave as autonomously as possible, in order to reduce the maintenance overhead and improve the usability of the security infrastructure.

\paragraph{Privacy} Besides malicious adversaries, it is also possible to find honest but curious adversaries. These adversaries are usually authorized entities (e.g. edge data centers, infrastructure providers) whose secondary goal is to know more about the entities that make use of their services. This knowledge can then be used in various ways: usage profiling, location tracking, disclosure of sensitive information, etc. All these adversaries represent a threat to the privacy of users. Unfortunately, all edge paradigms are open ecosystems, where multiple trust domains are controlled by different infrastructure owners. In such a context, it is not possible to know in advance if a certain service provider is trustworthy enough to respect the users' privacy. Therefore, this is a very serious threat that must be carefully considered.

There are various challenges in this area. First, personal data will be stored and processed by entities that are outside the control of the users. Therefore, it is essential to provide users with various efficient mechanisms that not only protect their information, but also allow users to query it and process it (e.g. auditable data, controlled disclosure). Second, it is necessary to achieve a balance between anonymity and responsibility. In this dynamic environment, users have the right to protect their identity and their personal data, but also have the responsibility to behave honestly. If a user misbehaves, it should be possible to use some mechanisms to identify the malicious party. Finally, we need to consider that human mobility is, in fact, quite predictable (cf.~\cite{Song1018}): we usually go to the same places, follow the same routine every day. As a result, users will probably make use of the same edge data centers over and over. This poses a challenge to the development of privacy mechanisms that aim to protect the users' location and service usage.

\paragraph{Virtualization} The virtualization infrastructure is one of the core elements of edge paradigms, thus it is essential to protect it by designing and deploying security mechanisms in all edge data centers. Without these mechanisms, not only malicious insiders can take control of virtual machines deployed by users, but also malicious virtual machines can manipulate the services of edge data centers. There are numerous countermeasures that can be implemented in all commodity servers, such as isolation policies, hypervisor hardening, separation of roles and VMs, networking abstractions, and many others~\cite{Pek2013}. Note, however, that any mechanisms that depend on the restriction of physical access might be difficult to implement in this context.

\paragraph{Fault tolerance and resilience} No paradigm is ever going to be 100\% secure and immune from threats, and edge paradigms are no exception. Misconfigurations, vulnerabilities, outdated software, and other weaknesses will allow malicious adversaries to disable or take control of certain elements of the whole infrastructure. It is then necessary to integrate various mechanisms and strategies (e.g. redundant operations, failover capabilities, disaster recovery mechanisms) that will allow the service infrastructure to continue its intended operation. However, the deployment of the edge data centers at the edge of the network is a double-edged sword. On the one hand, protection mechanisms can take advantage of the fact that various infrastructure providers might be available at the same location. On the other hand, as services are provided at a local level, there might be situations where no replacement is available.

\paragraph{Forensics} As we have already mentioned, no matter what protection mechanisms are put in place, edge paradigms will be successfully attacked. These attacks will leave certain evidence behind, which can be used to reveal information about the attacker and his methods. The goal of forensics is to identify, recover, and preserve this evidence, so it can be presented in court. The management of evidence in edge paradigms is a very complex issue, mainly due to the existence of multiple actors, infrastructures, technologies, and scenarios. Nevertheless, it might be possible to make use of existing research in related areas, such as cloud forensics (cf.~\cite{Quang15,VirtForensics10,Rahman16}), to solve certain issues such as mobile forensics, virtualization forensics, and storage forensics.

Besides, Wang et al.~\cite{WangF15} and Zawoad et al.~\cite{ZawoadF15} have provided a detailed analysis of the main requirements of fog computing forensics and mobile cloud computing forensics, respectively. Both works agree that there are various common challenges in this area, such as i) storing trusted evidence in a distributed ecosystem with multiple trust domains, ii) respecting the privacy of other tenants when acquiring and managing evidence, and iii) preserving the chain of custody of the evidence. Then again, both works agree that edge data centers should need less computational resources to manage potential evidence: due to their geographical location and their local scope, they do not manage as many resources (e.g. network traffic, virtual machines) as centralized cloud infrastructures.


\section{Security Challenges and Opportunities}
\label{sec_cha}

In the previous sections, we have reviewed the similarities and differences between all edge paradigms, and we have provided a detailed analysis on the threats that can target these paradigms -- and the security mechanisms that should be used to protect them. In this section we will provide an analysis of the state of the art regarding security in all edge paradigms (section~\ref{sec_cha::cha}), and we will conclude such analysis with a discussion on existing shortcomings and potential research areas (section~\ref{sec_cha::summ}). As with Section~\ref{feat_syn::chasyn}, we will point out in our analysis potential synergies between all edge paradigms. Note, however, that we will also consider in our analysis other related paradigms (e.g. cloud computing, grid computing, peer-to-peer computing) and some of the enabling technologies that are used by edge paradigms (e.g. wireless networks, distributed and peer-to-peer systems, virtualization platforms~\cite{Vaquero14}).

There are several reasons for this. Some of the underlying assumptions of the security mechanisms that were designed for related paradigms do not conflict with the requirements of edge paradigms. For example, certain peer-to-peer security mechanisms only require a decentralized infrastructure of peers that can communicate with each other. Other security protocols are independent of the underlying technologies that implement them, thus they can be easily adapted to other environments. Moreover, some security mechanisms were designed with a specific scenario in mind, but their functional elements can easily be mapped to edge paradigm scenarios. For example, the security components of certain trust management systems for grid computing only assume that servers hosted at different administrative domains can exchange trust information securely. These building blocks can be mapped to a edge paradigm scenario where multiple edge data centers that belong to different trust domains exchange information in a secure way. It is obvious that all these security mechanisms should not be adapted without an extensive analysis, yet they can prove that researchers do not need to start from scratch when designing security mechanisms for edge paradigms.

\subsection{Specific challenges and promising solutions}
\label{sec_cha::cha}

\subsubsection{Identity and Authentication}

At present, there are no research works that analyze how to identify and authenticate the members of a world-wide infrastructure of interconnected edge data centers owned by different companies and individuals. Yet it might be possible to look for the solution to this problem in other related fields, such as federated cloud computing and peer-to-peer computing. In fact, there are multiple approaches that pursue the creation of inter-cloud identity management systems~\cite{Toosi14}. Such approaches make use of various standards, like SAML and OpenID, in order to provide Single-Sign On (SSO) authentication between clouds. As for peer-to-peer computing, there are also several mechanisms that provide mutual authentication without having to connect to a central authentication server~\cite{Touceda15}. As the design of these approaches is compatible with the underlying infrastructures of edge paradigms, all these approaches might be adapted to handle the authentication of edge data centers that belong to different trust domains.

On the other hand, there are some authentication infrastructures, which focus on user authentication within the same trust domain, explicitly designed for edge paradigms. For example, Donald et al.~\cite{Donald15} defined a centralized infrastructure for MCC where a single trusted third party serves as the authentication server. However, this approach requires the authentication server to be accessible at all times, thus their applicability is limited. In another work, Ibrahim~\cite{Ibrahim16} developed a user authentication system that allows any fog user and fog node to mutually authenticate each other, yet this approach forces all fog nodes to store certain credential information of all the users of the trust domain. There are other works in the areas of MCC~\cite{Tsai15} and fog computing~\cite{StojmenovicFSec05} that are able to authenticate users, even if the authentication server is not reachable, with less overhead. This is achieved by using pairing cryptosystems and secure hardware~\cite{Tsai15}, or by using hybrid encryption (public-key and symmetric-key encryption)~\cite{StojmenovicFSec05}. Although these mechanisms focus on user authentication, they might be useful for authenticating a federation of edge data centers that belong to the same trust domain.

Precisely, on the subject of user authentication, as edge data centers are located in the vicinity of end-users, researchers have proposed various authentication schemes that make use of location-specific information. For example, in the context of federated mobile cloud computing, Shouhuai et al.~\cite{Shouhuai16} introduced the concept of situational authentication, which is based on notions such as ``whom you are with'', ``where you are'', and ``what time is it''. Other authors, such as Bouzefrane~\cite{Bouzefrane14}, use Near Field Communication (NFC) to verify that a mobile device is offloading tasks to an authorized local cloudlet. Note that the notion of location-based authentication has been already studied in several other fields (e.g. wireless sensor networks~\cite{Zeng2010}, Internet of Things~\cite{Habib15}), providing various mechanisms that could be adapted to edge paradigms.

As for user mobility, there have been some protocols that have tried to implement a secure and efficient handover authentication in MCC scenarios. For example, Yang et al.~\cite{Yang2015Hand} provided an efficient design that allowed a mobile client to migrate from one region to another. Note that these protocols usually need to access an authentication server in a centralized cloud infrastructure, thus there is room for improvement. Finally, note that certain edge paradigms allow users to deploy their own personal data centers. Consequently, some works, such as the OPENi framework~\cite{McCarthy15}, have studied how to grant access to external users in such personal cloudlet platforms. In OPENi, the authentication component makes use of the OpenID Connect authentication layer, amongst other mechanisms. Therefore, the owner of the cloudlet decides which authentication servers he trusts, and what users are allowed to access the resources of the cloudlet.

\subsubsection{Access Control Systems}

There are very few studies that have investigated the development of fine-grained access control mechanisms in the context of edge paradigms. One example is the OPENi framework~\cite{McCarthy15}, which was mentioned in the previous section. In this framework, the authorization is based on OAuth 2.0, and the owner of the cloudlet defines the access rights of every resource by creating and storing access control lists (ACL) in a NoSQL database. This approach is more suitable for personal cloudlets, where their owners can define what operations can be performed on a resource by a certain user. Other approaches, like the one introduced by Huang et al.~\cite{HuangX14} and used by Stojmenovic et al.~\cite{StojmenovicFSec05}, use cryptographic primitives such as attribute-based encryption (ABE) to implement attribute-based access control policies. In this approach, users are provided with certain attributes, and access control rules connect such attributes with the operations that can be performed on a resource. This mechanism can be appropriate for a single trust domain, where service providers can use these attributes (alongside with their credentials) to get permission to deploy VMs in an edge data center.

Other authors have explored the deployment of policy enforcement components and the management of security policies. One simple example can be found in the architecture defined by Vassilakis et al.~\cite{Vassilakis2016a}, which made use of a formal methodology to deploy security components in MEC small cells. These components provide protection and access control to various MEC services, such as radio resources and virtualization services. However, one of the most prominent examples is the policy management framework for fog computing designed by Dsouza et al.~\cite{Dsouza14}. In this framework, the orchestration layer of the fog architecture is supported by a policy management module, which defines various components -- including a repository of rules, an attribute database, and a session administrator. Moreover, the policies can be enforced at various levels, such as edge data centers, VM instances, and IoT devices. This policy management framework does not have any special architectural requirements beyond the existence of a core infrastructure, thus it can be applied to other paradigms such as mobile edge computing.

As for the existence of federated and distributed access control architectures in other related paradigms, there is actually an extensive literature on this subject~\cite{Li15::Authe,Elsayed15}. Several of these mechanisms might be adapted to our context in order to solve existing open issues. For example, Almutairi et al.~\cite{Almutairi12} developed a distributed access control architecture for multicloud environments, based on role-based access control (RBAC) policies, that also provided inter-domain role mapping and constraint verification. This approach might be used to connect various entities that belong to different trust domains. Besides, there are other security mechanisms that, although not created for edge paradigms, might be suitable for certain scenarios. For example, the Direct Anonymous Attestation with Attributes (DAA-A) protocols allow anonymous users to prove that they possess a certain set of trusted attributes. These protocols can be implemented using the primitives defined in the Trusted Platform Module 2.0 (TPM 2.0) specification~\cite{Liqun15}, thus they can be applied to scenarios where two edge data centers need to prove that they have certain attributes (e.g. location, capabilities) without disclosing their owners.

\subsubsection{Protocol and Network Security}

All the communication technologies that are used by edge paradigms are either mature standards (e.g. TCP/IP stack, Wi-Fi) or are being extensively studied by both industry and academia (e.g. 5G, Sigfox). They define their own security protocols and mechanisms, which are able to provide privacy and data integrity between two authenticated entities. One of the challenges in this area is the distribution of the credentials that will be used to negotiate the session keys. Yet solutions are available, even if more research is needed. For example, a designated certification authority controlled by one infrastructure provider can distribute credentials to all the elements located within his trust domain. Cryptographic attributes, such as the attributes used in~\cite{HuangX14}, can also be used as credentials in order to exchange session keys~\cite{Gorantla2010}. Besides, there are several works in various areas, such as federated content networks~\cite{Pimentel201547}, that define how multiple trust domains can negotiate and maintain the interdomain credentials that will be used to establish secure channels. The requirements of these solutions are not very restrictive, thus they might be applied to edge paradigm environments.

Another aspect that we have to take into account is the security of the virtualized network infrastructure; that is, the network infrastructure that is used by the VMs deployed at edge data centers. As already pointed out by Ahmad et al. in their detailed analysis of the subject~\cite{AhmadSDN15}, both software defined networking (SDN) and network function virtualization (NFV) can be extremely useful in the context of edge paradigms. These approaches can be used in various ways, such as isolating different types of traffic even under adversarial conditions, isolating unsecure network devices, directing the traffic towards security devices, reconfiguring the systems in real time, etc. Notice that the original goal of NFV and SDN is to simplify the management of the network, by virtualizing the router functions and by implementing programmable network control and operation logic. These services are also beneficial for edge paradigms, as one of the challenges that need to be solved is the management of the network infrastructure~\cite{Kreutz15,YongNFV15}. Note, however, that both SDN and NFV have their own security challenges that need to be addressed~\cite{AhmadSDN15,Chen2016}.

\subsubsection{Trust Management}

Although trust is one of the most important security requirements in edge paradigms, the amount of research that has been conducted in this area as of 2016 is quite limited. Actually, most of the research has focused only on the area of mobile cloud computing, analyzing the trust relationships between users. For example, Petri et al.~\cite{Petri12} studied how various nodes could create a trustworthy peer-to-peer cloud, where feedback aggregation was used to identify egoist users. Also, Chen et al.~\cite{ChenT13} analyzed how call patterns can be used to derive the trust relationships between human users. One of the only works that explicitly analyzed how to calculate the reputation of edge data centers was developed by Hussain et al.~\cite{HussainT14}. In this work, the researchers describe the implementation of a centralized trust manager, which stores the reputation of LTE-deployed cloudlets. Using this system, users can rate the services of cloudlets anonymously.

Trust management has been a very active area of research in many other related fields~\cite{Shang12,Yan2014120,Corradini15}. Therefore, as with the other security properties, researchers might benefit from studying and adapting existing trust management systems. There are, in fact, several systems that might be applicable to edge paradigms, due to their focus on decentralized deployments and cross-domain relationships. One example is the self-managed trust management system by Kantert et al.~\cite{Kantert15}. In this work, autonomous servers from different administrative domains share their resources in a grid-like scenario. In contrast to other grid deployments, it is assumed that egoist or malicious servers will exist. Therefore, it is necessary to calculate a set of trust metrics in an autonomous and distributed way. This work might be used as a foundation for calculating trust values between edge data centers.

Another example is the quantitative trust management component, defined by Figueroa et al. and integrated into the Safety On Untrusted Network Devices (SOUND) platform~\cite{FigueroaUS15}. This platform is comprised by several communities of trust, which contains various hosts. Whenever two hosts from different communities interact, they take into account not only their mutual trust, but also the trust between their communities, and the trust between the community and the other host. Due to the similarities between the communities of trust and the edge trusted domains, the design of this particular component might be used as an input for the design of trust management systems deployed in edge data centers. Finally, Bennani et al.~\cite{Bennani14} defined a Bayesian network-based trust model for hybrid cloud computing environments. In this scenario, a private cloud can assess and track the reputation of various services provided by public clouds. This kind of approach might be used to track the reputation of services that are available to the whole ecosystem, such as Security-as-a-Service solutions.

\subsubsection{Intrusion Detection Systems}

Most of the research on the area of intrusion detection and prevention systems has focused on mobile cloud computing, with only a few exceptions such as the active honeypot system designed by Mtibaa et al~\cite{Mtibaa15} that focused on detecting local adversaries in mobile edge computing deployments. Yet some of these MCC-centric research works might be used for other paradigms, too. Gai et al.~\cite{Gai15} proposed a framework where mobile devices using 5G networks could delegate their intrusion detection tasks to centralized services located in the cloud. While this research was focused on centralized cloud services, it might be possible to adapt this framework to a more distributed approach, where the IDS services will be deployed in nearly located edge data centers. Such services will then have a comprehensive view of the state of their surroundings. Also, Shi et al.~\cite{Shi15} presented a distributed IDS deployed in a cloudlet mesh architecture. In this architecture, the members of the cloudlet can collaborate with each other and with external entities in order to detect malware, malicious attacks, and others. This type of collaborative IDS might also be used by a federation of edge data centers to monitor the traffic of a certain geographical location.

Although there is still work to be done, it is perfectly possible to reuse various IDS mechanisms and solutions developed for cloud computing~\cite{Iqbal201698} and other related paradigms. The reason is simple. The main task of edge data centers is to provide cloud computing capabilities to users. Therefore, edge data centers can benefit from IDS that monitor the behaviour of VMs, the internal network and their surroundings. The main challenge here is to deal with the distributed nature of the whole infrastructure, where multiple trust domains coexist. Yet many IDS solutions do not need of centralized infrastructures -- they can monitor their environment autonomously. One example is the CROW solution, developed by Pitropakis et al.~\cite{Pitropakis15}. This IDS solution makes use of the computational power of GPU cards to effectively monitor the health of each VMs, detecting both attacks against the infrastructure and the presence of malicious insiders. Other examples are the IDS solutions that rely on software based networking (SDN) principles, and provide services such as deep packet inspection, network reconfiguration, policy management, and flow-based anomaly detection, amongst others (cf.~\cite{AhmadSDN15}).

Moreover, there are actually various IDS frameworks whose goal is to interconnect and monitor different trust domains. Elements of these frameworks might be reused or adapted to our context. For example, Luo et al.~\cite{LuoFIDS13} introduced a security architecture for federated cloud environments that facilitates the early detection of cyberattacks and the deployment of early warning systems such as honeypots. Instances of this architecture need to be deployed in the centralized command and control center of every trust domain, thus its applicability to a N-tiered hierarchy needs to be further studied. Yet the architecture also introduces various mechanisms that allow multiple trust domains to coordinate in-cloud and cross-cloud defense activities. Finally, some studies, like~\cite{Encina14}, have provided an analysis of attacks that specifically target federated cloud environments -- and that can also target edge paradigms.

\subsubsection{Privacy}

In the field of edge paradigms, privacy is one area that has been particularly active in the last years. In fact, many of the security protocols presented in the previous sections (e.g. entity authentication~\cite{Tsai15} and authorization~\cite{StojmenovicFSec05,Liqun15}, trust management~\cite{HussainT14}) allow users to interact with edge data centers and other entities in an anonymous way. Besides, there is a multitude of data privacy mechanisms specifically developed for the mobile cloud computing paradigm. These mechanisms tackle several challenges such as enforcing privacy policies when migrating code and data amongst collaborating mobile devices~\cite{Ravichandran15}, and concealing the location of a set of clients that are located in the same geographical area by means of establishing a peer-to-peer network~\cite{ZhangMCC15}. These mechanisms are designed for a collaborative cloud of local devices, yet they only require that all devices are interconnected and know their physical location. Therefore, they might also provide some inputs on the design of future privacy mechanisms for collaborative edge data centers. Note that there are other mechanisms, such as the software-defined pseudonym system for vehicular networks developed by Huang et al.~\cite{HuangVN16}, that make full use of the concept of interconnected local cloudlets.

Moreover, privacy has been one of the most researched fields in cloud computing~\cite{CruzSP15}. There are various cloud computing processes that have been enhanced with privacy features, such as protecting VMs during their storage and execution, and migrating VMs from one data center to another~\cite{Ali15}. Most of these solutions do not need a centralized infrastructure, and only require of a Trusted Platform Module (TPM), thus they can be implemented in the commodity servers that are available in edge data centers. Moreover, there are specific privacy mechanisms, such as data encryption, secure data sharing, encrypted data search, integrity verification, and many others~\cite{Sun15}, whose main goal is to protect the personal data of users. Some of these mechanisms do not have high computational requirements, thus they can be implemented in the user devices that interact with the edge data centers.

Notice that there are some use cases (e.g. personal cloudlets, corporate environments) where there is a trust relationship between the users and the edge data centers located at their vicinity. In such cases, it is possible to deploy privacy helper entities in the edge data centers. These entities will act as a front-end for the users, and can implement various data privacy mechanisms. These mechanisms can be used to control the quality and granularity of the personal information that is received by service providers or other remote entities (cf.~\cite{Seneviratne13,Page14}). In addition, the privacy helpers can implement other privacy services, such as protecting the users' identities from other remote services by creating pseudonyms and/or concealing their addresses (cf.~\cite{Hassan15}). Finally, it should be noted that the edge paradigms themselves can actually be used to strengthen the privacy features of certain services, such as crowdsourcing. For example, Abdo et al.~\cite{Abdo15} demonstrated that, by deploying a crowdsourcing platform in a trusted edge data center, it was possible to protect the anonymity of the participants of certain location-based services.

\subsubsection{Virtualization}

In the context of cloud computing, the security of virtualization infrastructures is a field that has been intensively studied in recent years~\cite{Lombardi2014}. Fortunately, many secure virtualization mechanisms do not need centralized managers or specific hardware unavailable to commodity servers. Therefore, they can be applied to the virtualization infrastructures that are used in edge paradigms. One clear example is the notion of Virtual Trusted Platform Modules (vTPM)~\cite{Berger06}. By virtualizing Trusted Platform Modules (TPM), vTPMs are able to provide TPM services (e.g. secure storage, cryptographic functions) to any virtual machine that is running on top of a hypervisor. In fact, existing hypervisor platforms, such as Xen and Hyper-V, already provide support for vTPMs. These services have been used to implement various security services that are relevant to edge paradigms, such as VM creation and cloning~\cite{MaVTPM12}, VM migration~\cite{Mahdi14}, platform attestation~\cite{Jacquin2015}, and many others (data storage, secure rollbacks).

Besides, in the area of mobile cloud computing, there are some research studies that propose secure computation offloading solutions. For example, Hao et al.~\cite{Hao15} proposed a system that allows a subset of a mobile application to securely run in a cloud server. Also, Dhanya et al.~\cite{Dhanya15} proposed a secure partitioning mechanism that kept the most sensitive or vulnerable parts of an application in the mobile device. These solutions might be adapted to other edge paradigms, as there might be some cases where a VM only needs to send a small agent to other data centers (cf.~\cite{Borcea15}).

\subsection{Summary}
\label{sec_cha::summ}

\begin{table*}[!htb]
\centering
\begin{scriptsize}
\begin{tabular}{|c|c|c|c|c|}
  \hline
   & \textit{Fog Computing} & \textit{MEC} & \textit{MCC} & \textit{Other paradigms} \\
  \hline
  \textit{Identity and Authentication} &
  \cite{StojmenovicFSec05}\cite{Ibrahim16} &
  --- &
  \makecell{\cite{Donald15}\cite{Tsai15}\cite{Shouhuai16} \\ \cite{Bouzefrane14}\cite{Yang2015Hand}\cite{McCarthy15}} &
  \makecell{\cite{Toosi14}\cite{Touceda15}\cite{Zeng2010} \\ \cite{Habib15}} \\
  \hline
  \textit{Access Control Systems} &
  \cite{HuangX14}\cite{StojmenovicFSec05}\cite{Dsouza14} &
  \cite{Vassilakis2016a} &
  \cite{McCarthy15} &
  \makecell{\cite{Li15::Authe}\cite{Elsayed15}\cite{Almutairi12} \\ \cite{Liqun15}}\\
  \hline
  \textit{Protocol and Network Security} &
  --- &
  --- &
  --- &
  \makecell{\cite{HuangX14}\cite{Gorantla2010}\cite{Pimentel201547} \\ \cite{AhmadSDN15}\cite{Kreutz15}\cite{YongNFV15}}\\
  \hline
  \textit{Trust Management} &
  --- &
  --- &
  \cite{Petri12}\cite{ChenT13}\cite{HussainT14} &
  \makecell{\cite{Shang12}\cite{Yan2014120}\cite{Corradini15} \\ \cite{Kantert15}\cite{FigueroaUS15}\cite{Bennani14}}\\
  \hline
  \textit{Intrusion Detection Systems} &
  --- &
  \cite{Mtibaa15} &
  \cite{Gai15}\cite{Shi15} &
  \makecell{\cite{Pitropakis15}\cite{AhmadSDN15}\cite{LuoFIDS13} \\ \cite{Encina14}}\\
  \hline
  \textit{Privacy} &
  \cite{StojmenovicFSec05} &
  --- &
  \makecell{\cite{Tsai15}\cite{HussainT14}\cite{Ravichandran15} \\ \cite{ZhangMCC15}\cite{Seneviratne13}\cite{Hassan15} \\ \cite{Abdo15}\cite{HuangVN16}}&
  \makecell{\cite{Liqun15}\cite{CruzSP15}\cite{Ali15} \\ \cite{Sun15}\cite{Page14}}\\
  \hline
  \textit{Virtualization} &
  --- &
  --- &
  \cite{Hao15}\cite{Dhanya15}\cite{Borcea15} &
  \cite{Lombardi2014} \\
  \hline
  \textit{Forensics} &
  \cite{WangF15} &
  --- &
  \cite{ZawoadF15} &
  \cite{Pichan15} \\
  \hline
\end{tabular}
\end{scriptsize}
\caption{State of the art in Edge security as of Q1 2016}\label{T:SotA}
\end{table*}

Table~\ref{T:SotA} provides a summary of the state of the art that was reviewed in the previous section. In this table, all studies are classified according to the original paradigm for which they were designed. One obvious conclusion is that there are very few studies that have been specifically designed for fog computing and mobile edge computing, compared to the amount of studies that have been focused on mobile cloud computing. The reasons are simple: i) these paradigms were created very recently, and their infrastructure has not been fully defined, and ii) the mobile cloud computing paradigm has been studied longer. The reader should note, however, that many studies in the area of mobile cloud computing have not targeted the security of edge data centres (i.e. cloudlets), but distributed clusters of mobile devices instead.

Even if the number of studies that target edge paradigms is quite limited, it does not mean that researchers must start from zero when developing new security mechanisms. As we have seen in the last section, it might be possible to use the security mechanisms and components that have been designed for other related paradigms as a foundation for the development of novel edge security mechanisms. Moreover, we also have shown in the most recent section that it might possible to reuse or adapt various security mechanisms that were specifically designed for one edge paradigm to the other edge paradigms. However, it is necessary to analyse how the specific nuances of every edge paradigm -– like underlying features of mobile network operator infrastructures or user-owned edge data centres –- will affect this adaptation process.

Having said this, several issues will need to be studied and evaluated in the near future. Some examples of these issues are explained here briefly. It is necessary to investigate the impact that certain attacks, such as denial of service, rogue data centres and malicious VMs, will have on the service infrastructure. In addition, it must be assessed how such attacks can be detected and neutralised by intrusion detection/prevention systems. The edge paradigm ecosystem must provide support for various identity management frameworks, including those used by prominent application scenarios like the Internet of Things. It must be possible for administrators to maintain a consistent network configuration and access control policy across all elements of the edge infrastructure with as little overhead as possible. Besides, we need to analyse how trust management systems can benefit other security mechanisms as well as the exact impact that edge paradigms will have on the privacy of their users. It is essential to reduce the latency of all security mechanisms as much as possible, and to study the security of mobile entities in this context.

Furthermore, there are certain research areas that have been neglected in the context of edge paradigms, such as secure software engineering, security and usability, fault tolerance and resilience, and forensics. All of them are essential in this context. By considering the specific features of edge paradigms (e.g. context awareness or interaction with mobile clients) during the development of security-aware software systems, the vulnerabilities specific to our context will be greatly reduced. Usability is another essential factor, as the development of usable security mechanisms will limit misconfigurations and facilitate the maintenance of the whole ecosystem. Thanks to fault tolerance, the service infrastructure will be able to continue its operation, even if at a reduced level. Last, malicious adversaries can be identified and prosecuted if effective forensics procedures are in place.


\section{Conclusions}
\label{conclusions}

In this study, we have analysed from a holistic perspective the security threats and challenges that affect edge paradigms, such as fog computing, mobile edge computing, and mobile cloud computing. In the first part of our analysis, we identified the features and problems that are common to all edge paradigms. In the second part, we provided a novel analysis of the multiple threats that target all edge paradigms, alongside a detailed study regarding the state of the art of security mechanisms that should be integrated into all edge paradigms. As a conclusion of this analysis, we have shown that research should not be compartmentalised, but all edge paradigms should consider the advances in other paradigms. Nevertheless, the security of edge paradigms is still in its infancy; thus, there are multiple open issues that merit consideration in the near future.


\section*{Acknowledgements}

This work was partially supported by the Spanish Ministry of Economy and Competitiveness through the PERSIST (TIN2013-41739-R) project, and by the European Commission through the NeCS (H2020-MSCA-ITN-2015-675320) project, which is under the umbrella of the Marie Sklodowska-Curie Innovative Training Networks (ITN).




\section*{References}

\bibliographystyle{elsarticle-num}
\bibliography{bibtex}

\end{document}